# Chocs technologiques, chocs des prix et fluctuations du chômage en République Démocratique du Congo

Antoine KAMIANTAKO MIYAMUENI[1]

Henry NGONGO MUGANZA[2]

Abstract

The main purpose of this research is to analyze the effects of macroeconomic shocks on unemployment fluctuations in the Democratic Republic of Congo (DRC). Using the SVECM model on DRC data for the period 1960 to 2014, the conclusion is that the high and persistent level of unemployment is mainly explained permanently by technological and price shocks.

*Résumé*

L'objet de cet article est d'analyser les effets des chocs macro-économiques sur les fluctuations du chômage en République Démocratique du Congo (RDC). En utilisant le modèle SVECM sur les données de la RDC pour la période allant de 1960 à 2014, on aboutit à la conclusion selon laquelle le niveau élevé et persistent du chômage est, à long terme, essentiellement expliqué de manière permanente par les chocs technologiques et des prix.

Mots-clés : Chocs macro-économiques, chocs technologiques, chocs es prix, chômage, SVECM

---

[1] Professeur des Universités, faculté des Sciences Économiques et de Gestion (UNIKIN), faculté d'Administration des Affaires et Sciences Économiques (UPC-Kinshasa), Tél. : +243999909613, Mail : kamiant@gmail.com

[2] Doctorant à l'école Doctorale de la FASE - UPC/Kinshasa et assistant d'enseignements à la Faculté des Sciences Économiques et de Gestion, Université Évangélique en Afrique (UEA - Bukavu). Tél. : +243995157604, Mail : henryngongo06@gmail.com; henryngongo@uea.ac.cd

## I. Introduction

Le débat sur le chômage a connu un regain d'intérêt dans les années 1980 à la suite du renouvellement théorique construit autour de l'analyse micro-fondée initiee par les travaux de Friedman (1968) et Phelps (1967). A ce jour, l'explication du chômage reste dominée par deux courants de pensées : les nouveaux classiques et les nouveaux keynésiens.

Selon les nouveaux classiques, notamment Lucas (1972, 1976), Barro (1974) et Sargent et Wallace (1975), les fluctuations cycliques sont des réponses optimales de l'économie à des chocs exogènes. Toute variation des prix, de l'offre ou de la demande résulte d'un processus permanent d'ajustement. Lorsque les salaires réels sont temporairement élevés, les ménages seraient prêts à travailler plus et vice-versa. Le marché du travail étant en équilibre de plein emploi, le chômage observé est le résultat d'un processus de substitution inter-temporelle. Il s'agit donc d'un chômage tout à fait volontaire.

Pour leurs part, les nouveaux keynésiens plaident pour la rigidité nominale et réelle des prix. Ils soulignent que les prix sont rigides parce qu'ils n'évoluent pas en fonction des modifications du marché (Phelps et Taylors, 1977). Aussi, les contrats étant négociés de manière échellonnée, il y a un décalage entre les variations de prix et l'adaptation des salaires (Taylors, 1980). De leur coté, Akerlof et Yellen (1985) font valoir qu'en situation de concurrence imparfaite, les ajustements des prix ne correspondent pas aux fondementaux du marché. Il en est de même pour le coût de menu qui empêche les entreprises à modifier leurs prix (Mankiw, 1985).

Par rapport aux rigidités réelles, Azariadis et Stiglitz (1983) montrent que l'aversion pour le risque des salariés est prise en charge par les employeurs de telle sorte que les entreprises ne modifient pas immédiatement les salaires réels en cas de choc économique. Pour Akerlof (1982), les employeurs garantissent un salaire élevé en contrepartie des efforts du salarié quelque soit la conjoncture. Les rigidités réelles peuvent également découler du pouvoir de négociation salariale des syndicats à en croire McDonald et Solow (1981). Les nouveaux keynésiens considèrent donc le chômage comme un phénomène involontaire.

En effet, l'un des cadres d'analyse priviligié par les nouveaux keynésiens pour analyser le chômage est le modèle « WS-PS » mis en oeuvre par Nickell, Jackman et Layard (1991) et développée par Pierre Cahuc (1999). Selon ce modèle, il existe une relation conjointe entre le chômage, les salaires et les prix. Pour obtenir le niveau du chômage d'équilibre, il suffit de confronter la courbe des salaires à celle de la détermination des prix. Le niveau de chômage d'équilibre déterminé ne peut être modifié que par les chocs structurels affectant les déterminants des salaires ou des prix, notamment les chocs pétroliers, les chocs sur le niveau des prélèvements directs ou indirects et les chocs de taux d'intérêt réels.

Sur le plan empirique, plusieurs études se sont penchées sur l'identification des chocs macro-économiques et évaluer leur importance sur le chômage beaucoup plus dans les pays développés que dans les pays en développement.

Fabiani, Locarno, Oneto et Sestito (2001) analyse les principales sources du chômage en Italie entre 1954 – 1998. A l'aide d'un modele VAR structurel sur les données trimestrielles, ils aboutissent à la conclusion selon laquelle le niveau du chômage Italien est attribué aux chocs de productivité et d'offre du travail.

Dans le même ordre d'idée, Hansen et Warne (2001) réalisent une étude sur les causes du chômage au Danemark pour la période 1905 – 1992. En utilisant le modèle à tendances communes avec vecteur de co-intégration, ils réalisent que ce sont les chocs d'offre qui expliquent principalement le chômage au Danemark. Dolado et Jimeno (1997) de leurs coté, analysent les principales causes du chômage en Espagne entre 1973 - 1995 à partir d'un modèle VAR structurel. Ils ont constaté que le chômage espagnol est expliqué à la fois par les chocs d'offre et de demande.

Jacobson, Vredin et Warne (1997) utilisent l'approche VARX pour analyser la persistance et les sources du chômage dans les pays scandinaves (Norvège, Danemark et Swede) de 1965 à 1989. Ils réalisent que le phénomène d'hystérèse dans ces pays est expliqué par le choc technologique, le choc d'offre du travail et le choc des salaires. Ce dernier est la seule source commune d'hystérèse dans les trois pays. Les chocs technologiques ont des effets permanents sur le chômage en Norvège, alors que pour le Danemark ce sont les chocs d'offre. Sunde et Akanbi (2016) pour leur part, ont recouru à un modèle VAR structurel pour analyser les sources du chômage en Namibie sur données annuelles pour la période 1980 – 2013. Ils aboutissent à la conclusion selon laquelle ce sont les chocs d'offre, de la demande, de salaire et des prix qui expliquent le chômage namibien. Il faut noter que les chocs des prix n'affectent le chômage qu'à long terme.

L'objet de cet article est d'expliquer les fluctuations du chômage en RDC durant la période allant de 1960 à 2014. Il est question d'évaluer l'ampleur des différents chocs macro-économiques sur le chômage congolais et identifier leur caractère permanent ou transitoire. Le choix porté sur la RDC tient au fait que celle-ci offre un champ très peu investi et dont l'analyse s'impose pour comprendre les fluctuations du chômage et mieux orienter les politiques publiques. Le taux de chômage est l'un des principaux indicateurs de la performance économique dont l'analyse serait très importante aujourd'hui en RDC pour au moins deux raisons :

1°) le niveau élevé et persistent du chômage qui caractèrise l'économie congolaise ;

2°) l'absence d'évidence empirique pour essayer de comprendre et expliquer ce phénomène et surtout, orienter l'action des pouvoirs publics.

C'est dans ce cadre que nous nous proposons de répondre à la question fondamentale ci-après : *quels sont les chocs qui expliquent le déséquilibre persistant sur le marché congolais du travail ?*

## 2. Méthodologie

Depuis la remise en cause des modèles structurels keynésiens par Lucas (1976) et Sims (1980), les modèles vectoriels et les modèles d'équilibre général occupent une place de choix dans les analyses empiriques pour leur capacité à analyser le rôle joué par différents chocs exogènes dans l'explication entre autres du chômage en vue de l'orientation des politiques économiues. Pour répondre à notre question de départ, nous allons faire recours au modèle VAR structurel qui est l'un des modèles dominants dans la littérature empirique.

Les données utilisées étant des séries chronologiques, nous commencerons par étudier les propriétes statistiques de manière aussi bien uni-variée que multivariée. Ces analyses porteront essentiellement sur le test de racine unitaire et celui de co-intégration. Pour ce faire, nous utiliserons le logiciel JMulTi4. Celui-ci présente l'avantage d'estimer la version structurelle du modèle VAR lorsque les données sont co-intégrées.

Le cadre de référence pour le reste de nos analyses va s'appuyer sur le modèle « WS-PS » tel que decrit par Layard, Nickell, et Jackman (1991).

S'inscrivant dans la logique de la nouvelle école keynésienne, ce modèle offre un cadre théorique simple et opérationnel à partir duquel les évolutions macro-économiques du chômage et des salaires sur une longue période peuvent être analysé. (Chelini, Prat, et al., 2013). Ce faisant, nous reprenons le cadre élaboré par Dolado et Jimeno (1997) et Brüggemann (2006) pour orienter notre réflexion au cas de l'économie congolaise.

### 3. Cadre théorique

Les équations structurelles du modèle théorique se présente sous la forme log-linéarisée comme suit :

$$y = \varphi(d - p) + a\vartheta \tag{1}$$
$$y = n + \vartheta \tag{2}$$
$$p = w - \vartheta + \psi \tag{3}$$

Les équations (1), (2) et (3) désignent respectivement la fonction de la demande agrégée où $\varphi > 0$, la fonction de production Cobb-Douglas et la règle de fixation des prix en concurrence imparfaite. Les variables $y$, p, $n$, $w$ et (d-p) désignent respectivement la production, les prix, l'emploi, le salaire et la composante exogene de la demande agrégée qui reflète la politique fiscale et monetaire. $\vartheta$ et $\psi$ sont les facteurs de changement de la productivité et de la fixation des prix; $d$ étant l'indice de la dépense nominale. Ce systeme d'équations représente le marché des biens. Suivant Dolado et Jimeno (1997), l'offre du travail peut être modélisée comme suit :

$$l = \alpha(w - p) - bu + \varsigma \tag{4}$$
$$w = w^* + \varepsilon_w + \gamma_1 \varepsilon_d + \gamma_2 \varepsilon_p \tag{5}$$
$$w^* = \arg\{n^e = \lambda l_{-1} + (1-\lambda)n_{-1}\} \tag{6}$$
$$u = l - n \tag{7}$$

avec $l$ : la population active, $n^e$ : le niveau attendu de l'emploi, $u$ : le taux de chômage, $\varsigma$ : les facteurs institutionnels ou démographiques susceptibles d'affecter l'offre du travail et $\varepsilon_w$, $\varepsilon_d$ et $\varepsilon_p$ sont respectivement les chocs de salaire, de la demande agrégée et de prix. L'équation (4) est la fonction d'offre du travail qui dépend des salaires réels $(w-p)$ et du taux de chômage $(u)$; avec $\alpha > 0$ et $b > 0$ reflétant la découragement dû au chômage de longue durée. L'équation (5) caractérise le mode de fixation des salaires. On admet que les négociations salariales sont partiellement indexées aux effets des chocs de demande et des prix par le biais des coefficients d'indexation $\gamma_1$ et $\gamma_2$ de sorte que, si $\gamma_i = 0$ $(i = 1,2)$, alors l'indexation n'est pas parfaite. L'équation (6) est la fonction de salaire et $w^*$ désigne le salaire nominal ciblé. On admet, en outre, que l'hystérèse est partielle lorsque $0 < \lambda < 1$ et totale si $\lambda = 0$. Partant, l'équation (6) peut se réécrire comme suit : $w^* = n^e = n_{-1}$. Il faut, par ailleurs, préciser que les chocs stochastiques régissant l'évolution des facteurs exogènes ($\varphi$, $\vartheta$, $\psi$ et $\varsigma$) suivent des processus aléatoires :

$$\Delta \varphi = \varepsilon_d \qquad (8)$$
$$\Delta \vartheta = \varepsilon_s \qquad (9)$$
$$\Delta \psi = \varepsilon_p \qquad (10)$$
$$\Delta \varsigma = \varepsilon_l \qquad (11)$$

avec $\varepsilon_d$, $\varepsilon_s$, $\varepsilon_p$ et $\varepsilon_l$ qui sont tous $\sim iid(0, \sigma^2)$. En résolvant les équations (1) – (11) par rapport au chômage, nous obtenons ce qui suit :

$$(1-\rho L)u = (1+b)^{-1} \begin{Bmatrix} -\varphi(1-\gamma_1)\varepsilon_d + \varphi\varepsilon_w + \varepsilon_l + [\varphi(1+\gamma_2)-\alpha]\varepsilon_p \\ -(\varphi+a+\alpha-1)\varepsilon_s \end{Bmatrix} \qquad (12)$$

où $L$ est l'opérateur de retard et $\rho = (1+b-\lambda)/(1+b)$. En cas d'hystérèse partielle, la persistance du chômage est une fonction croissante à la fois de l'effet de découragement $(b)$ et de l'influence de l'emploi décalé sur la détermination des salaires $(\lambda)$. Pour $b$ finie, $\rho = 1$ est équivalent à $\lambda = 0$ de sorte que l'hypothèse d'hystérèse totale suppose que le processus suivi par le chômage est intégré d'ordre 1.

Sur la base de l'hypothèse d'hystérèse total, les équations (1) – (11) permettent d'obtenir un système complet (un modèle réduit) dans lequel les variables peuvent être exprimées uniquement par des chocs structurels. On obtient alors après réarrangement le système suivant :

$$\Delta y = \varphi(1-\gamma_1)\varepsilon_d - \varphi(1+\gamma_2)\varepsilon_p - \varphi\varepsilon_w + (\varphi+a)\varepsilon_s \quad (13)$$

$$\Delta n = \varphi(1-\gamma_1)\varepsilon_d - \varphi(1+\gamma_2)\varepsilon_p - \varphi\varepsilon_w + (\varphi+a-1)\varepsilon_s \quad (14)$$

$$\Delta w = \varepsilon_w + \gamma_1\varepsilon_d + \gamma_2\varepsilon_p \quad (15)$$

$$\Delta p = \varepsilon_w + (1+\gamma_2)\varepsilon_p + \gamma_1\varepsilon_d - \varepsilon_s \quad (16)$$

$$\Delta u = (1+b)^{-1}\begin{cases} -\varphi(1-\gamma_1)\varepsilon_d + \varphi\varepsilon_w + \varepsilon_1 + \left[\varphi(1+\gamma_2)-\alpha\right]\varepsilon_p \\ -(\varphi+a+\alpha-1)\varepsilon_s \end{cases} \quad (17)$$

A partir des équations (13) – (17), il y a lieu de noter que le choc de la demande agrégée ($\varepsilon_d$) accroit le niveau de la production et de l'emploi. Il diminue par contre le chômage, le choc de salaire ($\varepsilon_w$) entraine la diminution de l'output et de l'emploi. Il augmente par ailleurs le salaire et le prix, les effets du choc de productivité ($\varepsilon_s$) dépendent de la valeur du paramètre $\varphi$. Si $\varphi > 1$, on assistera à une augmentation de l'output et de l'emploi alors que le chômage lui ne peut augmenter que si $\varphi < 1$.

On s'aperçoit que le choc de la demande agrégée ($\varepsilon_d$) a le même coefficient positif aussi bien dans l'équation d'emploi (14) que dans celle de la production (13). Il en est de même pour les chocs de salaire ($\varepsilon_w$) et de prix ($\varepsilon_p$). En plus, les chocs de la demande agrégée ($\varepsilon_d$) et de salaire ($\varepsilon_w$) ont les mêmes coefficients dans les equations de salaire (15) et de prix (16). De par ce constat, on peut soustraire l'emploi (14) dans la fonction de production (13) et le prix (16) de l'equation de salaire (15). Ce qui donne :

$$\Delta(y-n) = \varepsilon_s \quad (18)$$

$$\Delta(w-p) = \varepsilon_s - \varepsilon_p \quad (19)$$

$$\Delta p = -\varepsilon_s + (1+\gamma_2)\varepsilon_p + \varepsilon_w + \gamma_1\varepsilon_d \quad (20)$$

$$\Delta n = (\varphi + a - 1)\varepsilon_s - \varphi(1 + \gamma_2)\varepsilon_p - \varphi\varepsilon_w + \varphi(1 - \gamma_1)\varepsilon_d \qquad (21)$$

$$\Delta u = (1 + b)^{-1}\begin{Bmatrix} -(\varphi + a + \alpha - 1)\varepsilon_s + [\varphi(1 + \gamma_2) - \alpha]\varepsilon_p + \varphi\varepsilon_w \\ -\varphi(1 - \gamma_1)\varepsilon_d + \varepsilon_l \end{Bmatrix} \qquad (22)$$

Les équations (18) – (22) exploite les restrictions selon lesquelles les chocs de la demande $\varepsilon_d$ et de salaire $\varepsilon_w$ n'ont pas d'effet permanent sur la productivité $(y - n)$ et le salaire réel $(w - p)$. Seul le choc technologique $\varepsilon_s$ peut affecter le niveau de production à long terme en raison de l'hypothèse de rendement d'échelle constant. La composante permanente du salaire réel $(w - p)$ est quant à elle affectée par les chocs technologique $\varepsilon_s$ et des prix $\varepsilon_p$. Le choc d'offre du travail $\varepsilon_l$ par ailleurs n'a pas d'effet sur $y$, $n$, $w$ et $p$. En plus de supposer que le choc de demande $\varepsilon_d$ n'a pas d'effet sur le prix au cours de la période initiale, on retiendra aussi que le choc de prix $\varepsilon_p$ n'a pas d'effet permanent sur la productivité $(y - n)$.

### 4. Analyse empirique

Nous utilisons le modèle VAR structurel pour identifier les chocs transitoires et permanents qui expliquent la persistance du chômage en RDC pour la période allant de 1960 à 2014. Il s'agit des observations tirées de la base des données du Fonds Monétaire International (FMI) et subsidiairement complétée par celles de la Banque Centrale du Congo (BCC).

Avant toute analyse, il s'avère important de procéder par une analyse des propriétés statistiques des données. Cette analyse préliminaire des séries exige donc le recours au test de racine unitaire[3] et de co-intégration.

---

[3] Les resultats du test de stationarité de Dickey – Fuller Augmenté confirment que toutes les variables sont non stationnaires en niveau, mais stationnaire en différence première. Elles sont toutes intégrées d'ordre 1. Ce qui laisse présager l'existence d'un risque de co-intégration. Il faut dans ce cas procéder au test de co-intégration.

### 4.1. Test de co-intégration de Johansen

La 1ère étape de la procédure de Johansen consiste à déterminer le nombre de retards du système. Suivant le critère d'information de schwartz nous avons retenu un VAR avec $p = 1$. La deuxième étape consiste par la méthode de maximum de vraisemblance de tester l'existence d'une relation de long terme entre les variables et d'obtenir le nombre de vecteurs de co-intégration dans un cadre multivarié. Le principe de ce test est basé sur la comparaison du ratio de vraisemblance à la valeur critique. Ci-dessous le récapitulatif du test de co-intégration de Johansen.

Tableau n° 3. 1. Résultats du test de co-intégration

| Valeurs propres | $H_0$ | Test de la trace | | | Tests de S & L | | |
|---|---|---|---|---|---|---|---|
| | | Statistique | Valeurs critiques | | Statistique | Valeurs critiques | |
| | | | 5% | 1% | | 5% | 1% |
| 0.624 | $r = 0$ | 99.39 | 69,61 | 77,29 | 67,70 | 54,59 | 61,53 |
| 0.399 | $r \leq 1$ | 47.43 | 47,71 | 54,23 | 38,05 | 35,76 | 41,58 |
| 0.264 | $r \leq 2$ | 20.45 | 29,80 | 35,21 | 21,44 | 20,96 | 25,71 |
| 0.077 | $r \leq 3$ | 4.24 | 15,41 | 19,62 | 1,82 | 9,84 | 13,48 |

Sources : calculs effectués à partir des données avec le logiciel JMulTi4

Le résultat du test de Johansen révèle l'existence d'un seul vecteur de co-intégration entre le cinq variables parce qu'on obtient une valeur du test de trace (**99.39**) aussi bien que celui de Saikkonen et Lutkepohl (67,70) supérieure à leur valeurs critiques respectives (77,29) et (61,53) au seuil de 1%.

S'inscrivant dans la logique du modèle « WS – PS », la relation de co-intégration doit s'établir soit dans l'équation de demande de travail, soit dans celle de formation de salaire. Il convient dès lors de déterminer parmi ces deux équations celle dans laquelle le vecteur de co-intégration est identifié. Les resultats du test de restrictions linéaire montre le vecteur de co-intégration est identifié dans la relation de détermination de salaire conformément aux prescriptions du modèle théorique[4].

$$(w - p) = (y - n) - 0{,}433 u_t \qquad (23)$$

Il ressort de cette équation (23) que le salaire réel est positivement affecté par la productivité et négativement par le chômage. Ce qui est conforme à la théorie économique[5]. Les résultats de l'analyse préliminaire des données nous conduit à retenir un modèle VAR structurel avec vecteur de co-intégration. Ceci nous amène à présenter d'une part la structure du modèle SVECM et procéder à son identification, d'autre part. Ce n'est qu'après que les résultats de l'estimation pourrons être présentés.

### 4.2. Modèle vectoriel structurel à correction d'erreur (SVECM)

Pour identifier l'importance relative de chaque type de choc, il est usuel dans la littérature de recourir au modèle $VARs(p)$. L'objectif principal de cette méthodologie est de déterminer les réponses dynamiques des différntes variables économiques en combinant l'analyse des séries chronologiques et la théorie économique. Pour décomposer les chocs structurels en $r$ chocs transitoires et $n - r$ chocs permanents Pegan et Pesaran (2008) part d'un modèle SVAR à $n$ variables toutes intégrées d'ordre I qui peut être représenté comme suit :

$$A_0 X_t = B_1 X_{t-1} + \varepsilon_t \qquad (24)$$

En réécrivant l'équation (24) sous la forme VECM, on obtient le modèle qui suit :

---

[4]

[5] Par contre, s'agissant de demande de travail, on constate une relation positive entre salaire et productivité. Aussi, par rapport au chômage cette relation est négative. Ce qui n'est pas conforme avec la théorie économique.

$$\Delta X_t = -\alpha \beta' X_{t-1} + \psi \Delta X_{t-1} + \varepsilon_t \qquad (25)$$

A partir de l'équation (25), on déduit la forme structurel du modèle VECM (SVECM) dont la forme se présente comme suit :

$$A_0 \Delta X_t = -\alpha^* \beta^* X_{t-1} + B_2 \Delta X_{t-1} + \varepsilon_t \qquad (26)$$

### 4.3. Identification du modèle vectoriel structurel à correction d'erreur (SVECM)

Pour identifier exactement les chocs structurels, dans un système à cinq variables avec $r=1$ vecteur de co-intégration, on doit imposer dans la matrice d'impact à long terme ($\varepsilon B$) $k(k-r)/2 = 10$ restrictions de nature théorique linéairement indépendantes. Comme $r=1$, on aura alors un seul choc transitoire et $k^* = r(K-r) = 4$ chocs permanents. Conformément aux resultats du test de restrictions, le choc temporaire est ainsi associe à la variable choc sur le salaire. Ce qui revient à dire théoriquement que le chocs de salaire n'ont pas d'effets permanents sur les autres variables. Ce qui réduit le rang de la matrice $\varepsilon B$ de $k^*$ restrictions. Il faudra donc ajouter $k^*(k^*-1)/2 = 6$ restrictions afin d'identifier les chocs permanents et $r(r-1)/2 = 0$ restriction pour identifier le choc transitoire.

Au risque d'imposer les restrictions de façon aléatoire, nous recourons à celles qui sont de nature théorique. Ainsi, suivant l'hypothèse du rendement d'échelle constant ($\rho = 1$), le choc de demande, d'offre, de prix et de salaire n'ont pas d'effets permanents sur la productivité. Pour cette raison, les variables emploi, chômage, prix et salaire réel auront des coefficients nuls dans la première ligne de la matrice $\varepsilon B$. Il sera aussi supposé que le chômage n'a pas d'effets permanents sur la productivité, le salaire et le prix. En plus, la productivité n'a pas d'effets permanents sur le prix à long terme. A court terme par contre, l'emploi n'a aucun effet sur le salaire. A cet effet, les restrictions de long terme ($\varepsilon B$) et celles de court terme ($B$) dans la matrice d'impact peuvent être interprétées comme suit :

$$B = \begin{pmatrix} . & . & . & . & . \\ . & . & . & . & . \\ . & . & . & . & . \\ . & . & . & . & . \\ . & . & . & . & . \end{pmatrix} \quad \text{et} \quad \varepsilon B = \begin{pmatrix} . & . & 0 & . & 0 \\ 0 & . & 0 & 0 & 0 \\ . & . & 0 & 0 & 0 \\ . & . & 0 & . & . \\ . & . & 0 & . & . \end{pmatrix} \quad (29)$$

### 4.4. Résultats obtenus

En accord avec les restrictions imposées sur les matrices $\varepsilon B$ et $B$, les résultats de l'estimation de la version structurelle du modèle vectoriel à correction d'erreur (SVECM) qui est adapté pour identifier les chocs exogènes susceptibles d'affecter, de manière transitoire ou permanente, le chômage se présentent alors comme suit :

Tableau n° 3. 2. Matrice des coefficients d'impact de court terme ($B$)

|   | $\varepsilon^P$ | $\varepsilon^s$ | $\varepsilon^w$ | $\varepsilon^d$ | $\varepsilon^l$ |
|---|---|---|---|---|---|
| $p$ | 0,1797 (1,9903*) | 0,1133 (1,3011) | -0,3146 (-2,2421**) | -0,0868 (-09306) | 0,0231 (0,3611) |
| $(y-n)$ | -0,0861 (-0,9652) | 0,3153 (3,1975***) | 0,0715 (1,5004) | -0,1215 (-1,0980) | -0,1044 (-0,5898) |
| $(w-p)$ | 0,1224 (1,9609*) | 0,1031 (2,3106**) | 0,0422 (2,1476**) | -0,0572 (-1,4292) | -0,0049 (-0,1241) |
| $n$ | -0,1241 (-2,0304**) | -0,1177 (-2,6181**) | -0,0341 (-1,7448*) | 0,0291 (1,2484) | -0,0090 (-0,2984) |
| $u$ | 0,1255 (2,0512**) | 0,1168 (2,6102**) | 0,0339 (1,7459*) | -0,0258 (-1,1198) | 0,0016 (0,0461) |

Sources : Résultats estimés à partir du logiciel JMulTi 4
Notes : Le chiffre (.) désignent les valeurs de t-de student

Il ressort de la matrice d'impact de court terme que trois chocs ($\varepsilon^P$, $\varepsilon^s$ et $\varepsilon^w$) ont des effets significatifs dans les équations du chômage parmi lesquels deux ($\varepsilon^p$ et $\varepsilon^w$) présentent des signes attendus conformément au modèle théorique. Quant au choc technologique ($\varepsilon^s$), il présente un signe contraire à celui prescrit par le modèle. Ceci fait valoir que le choc technologique n'a pas entraîné la création de la richesse au cours de la période sous analyse. Par conséquent, le niveau d'emploi a baissé, d'où la hausse du chômage (cf. éq. 22). Cette baisse de l'emploi est expliquée par le signe négatif de l'effet du choc technologique sur l'emploi contrairement aux éxigences théoriques fournies par l'équation (21). On s'aperçoit aussi que le choc de salaire a un effet négatif sur le prix.

Tableau n° 3. 4. Matrice des coefficients d'impact de long terme ($\varepsilon B$)

|   | $\varepsilon^P$ | $\varepsilon^s$ | $\varepsilon^w$ | $\varepsilon^d$ | $\varepsilon^I$ |
|---|---|---|---|---|---|
| $p$ | -0,8197 (-1,6876) | 1,8869 (2,1973**) | 0 | 0,6277 (1,0052) | 0 |
| $(y-n)$ | 0 | 0,4189 (2,7580***) | 0 | 0 | 0 |
| $(w-p)$ | 0,1827 (2,0526**) | 0,0923 (1,1866) | 0 | 0 | 0 |
| $n$ | -0,1546 (-2,0092**) | -01873 (-1,9799*) | 0 | -0,0740 (-1,0119) | -0,0386 (-0,7891) |
| $u$ | 0,1544 (2,0179**) | 0,1875 (1,9851*) | 0 | 0,0821 (1,0136) | 0,0232 (0,7891) |

Sources : Résultats estimés à partir du logiciel JMulTi 4
Notes : Le chiffre (.) désignent les valeurs de t-de student

L'analyse des résultats fournis par la matrice d'impact de long terme place au premier plan le rôle joué par deux chocs permanents, à savoir le choc de prix ($\varepsilon^p$) et le choc technologique ($\varepsilon^s$), qui expliquent positivement et de manière significative la persistance du chômage. Comme à court terme, le choc technologique ($\varepsilon^s$) présente à long terme un signe non conforme au modèle théorique (cf. éq. 22). Ceci revient à dire qu'il a affecté négativement le niveau de production et via la loi d'okun la hausse du chômage. Ceci nous paraît normal dans un pays à forte croissance démographique où l'essentiel des biens et services consommés proviennent de l'extérieur. C'est cette défaillance du système productif interne qui justifie le signe négatif fourni par l'équation (21). Autrement dit, le choc technologique que connaît le pays est de nature à asphyxier la création d'emploi.

Par ailleurs, au niveau de l'équation (20), on constate qu'il affecte positivement le prix contrairement aux prescriptions théoriques. Nous osons croire que ce signe serait expliqué les effets de l'inflation importatée.

## 5. Intreprétation des résultats

Dans cette section il est question de présenter les fonctions de réponses du chômage aux impulsions des chocs retenus et discuter de l'importance de chaque choc transitoire ou permanent à partir de la décomposition de la variance.

### 5.1. Analyse des fonctions de réponses aux chocs

Les résultats des analyses des fonctions de réponses aux innovations sont décrits à l'aide des graphiques ci- dessous :

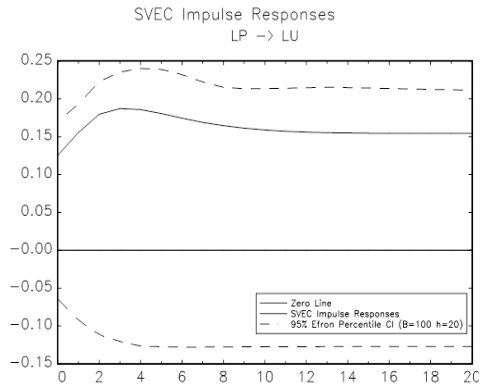
(a) réponse du chômage au choc des prix

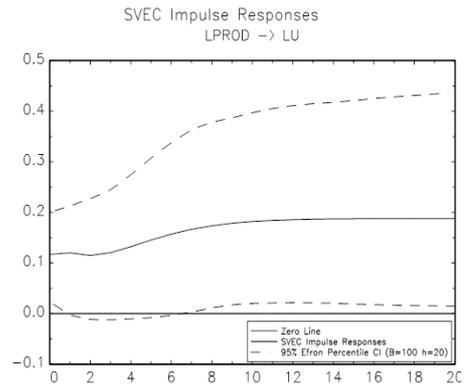
(b) réponse du chômage au choc

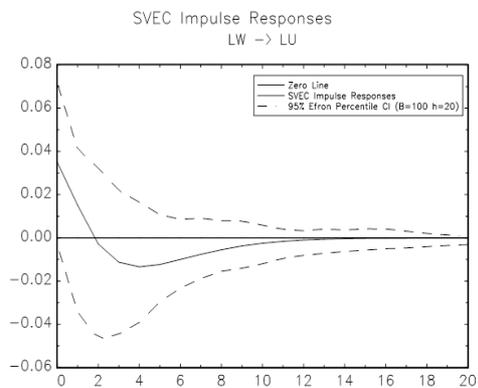
(c) réponse du chômage au choc de salaire

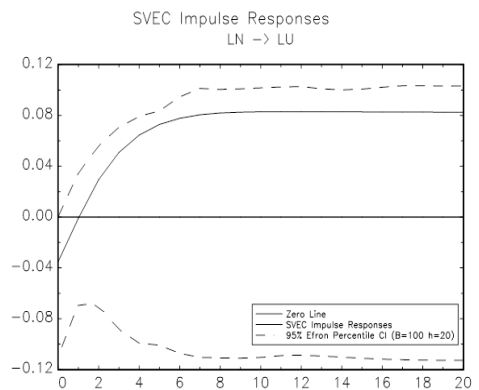
(d) réponse du chômage au choc de la

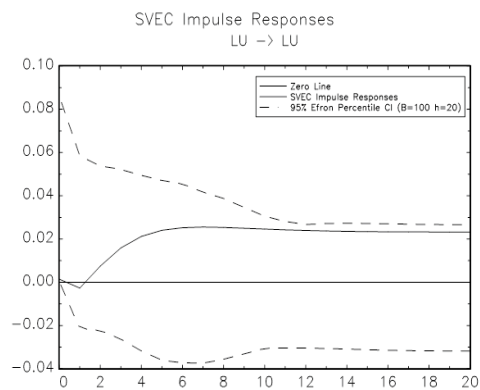
(e) réponse du chômage au choc d'offre

On constate que le choc de prix entraîne à la hausse le niveau du chômage. Cette tendance haussière est observable au cours de trois premières années. L'augmentation de prix s'est accompagnée par une élévation du chômage. A partir de la quatrième année, il devient stationnaire sur le reste de la période. Par rapport au choc technologique, on constate qu'il entraîne l'augmentation du chômage à partir de la quatrième année et devient stationnaire sur un horizon de long terme. Quant au choc de salaires, son effet négatif sur le chômage tend vers zéro jusqu'à disparaître à long terme. Ceci concorde avec les restrictions imposées dans le modèle théorique selon lequel le salaire n'a pas d'effets sur le chômage à long terme. Le choc de demande de travail entraîne la hausse du chômage sur un horizon de six ans. Au-delà de cette période, il devient stationnaire à long terme. Il en est de même pour le choc d'offre de travail à la seule différence qu'une baisse du chômage est observée à la première année. Ce choc a un effet positif sur le chômage à long terme.

### 5.2. Analyse de la décomposition de la varianc

Il est très intéressant d'analyser la décomposition de la variance pour apprécier le pouvoir de prédiction des différents chocs sur le chômage en RDC. Autrement dit, il s'agit de voir s'ils expliquent ou non l'erreur de prévision du chômage. Le résultat de cette analyse est présenté dans le tableau ci-dessous :

**Tableau n° 2.** Décomposition de la variance

| Périodes | Choc de prix $\varepsilon^P$ | Choc de technologique $\varepsilon^s$ | Choc des salaire $\varepsilon^w$ | Choc de la demande $\varepsilon^d$ | Choc d'offre $\varepsilon^l$ |
|---|---|---|---|---|---|
| 1 | 0,50 | 0,44 | 0,04 | 0,02 | 0,00 |
| 2 | 0,57 | 0,40 | 0,02 | 0,02 | 0,00 |
| 5 | 0,63 | 0,32 | 0,01 | 0,03 | 0,00 |
| 10 | 0,53 | 0,38 | 0,00 | 0,07 | 0,01 |
| 15 | 0,47 | 0,44 | 0,00 | 0,09 | 0,01 |
| 20 | 0,44 | 0,46 | 0,00 | 0,09 | 0,01 |

**Sources** : Nos calculs à partir du logiciel JMulTi 4

Il ressort de cette analyse que les chocs de prix ($\varepsilon^p$) et technologique ($\varepsilon^s$) expliquent respectivement 50 % et 44 % les fluctuations du chômage à court terme. Par contre, à long terme, le chômage est dû à 46 % au choc technologique et à 44 % au choc de prix. Pour ce qui est du choc de salaire ($\varepsilon^w$), il explique les fluctuations du chômage à hauteur de 4 % et cela seulement à court terme. A long terme, il n'a aucun effet. Quant au choc de demande ($\varepsilon^d$), il explique 2 % à court terme et 9 % à long terme les fluctuations du chômage. Ces dernières sont expliquées par le choc d'offre ($\varepsilon^l$) à hauteur de 1 % seulement à long terme.

## 6. Conclusion

Dans cet article nous avons cherché à évaluer le rôle joué par les chocs macro-économiques dans l'explication du chômage en RDC. A l'aide du modèle SVECM, nous obtenons les resultats selon lesquels au Congo le niveau élevé et persistent du chômage est essentiellement expliquée par le chocs technologique et de prix. Cette conclusion est validée par les résultats de l'analyse de la décomposition de variance de l'erreur de prévision.

Au vu de ces résultats, il y a lieu de noter que le chômage au Congo n'est pas un phénomène conjocturel, mais bien structurel. Cette observation ressort du fait les chocs de la demande aggrégée et celui de l'offre du travail ne semble pas pertinent au Congo. Comme ce sont les facteurs structurel qui déterminent le niveau élevé et persistent du chomage et bien la politique visa à favoriser la demande ne semble pas approprier pour le resorber.

Nous estimons qu'une extension du modèle thérioque retenu au cas d'une petite économie ouverte facilitera davantage la compréhension du chômage dans les économies en développement. Ceci permettre d'apprécier le rôle des chocs externes sur le chômage en exploitant la méthodologie

VAR structurel baysien. Cette approche semble efficace en comparaison avec celle exploitée dans le présent papier.

# Bibliographie